\documentclass[showpacs,aps,twocolumn]{revtex4}
\usepackage{epsfig}
\usepackage{graphicx}
\usepackage{amsmath,amssymb,amsfonts}
\usepackage{array}
\usepackage{url}
\usepackage[dvipsnames]{xcolor}
\usepackage{multirow}
\usepackage{float}
\usepackage{subcaption}
\usepackage{lineno}
\usepackage{xspace}
\usepackage{ulem}

\newcommand{\bef}{\begin{figure}}
\newcommand{\eef}{\end{figure}}
\newcommand{\bc}{\begin{center}}
\newcommand{\ec}{\end{center}}

\newcommand{\be}{\begin{equation}}
\newcommand{\ee}{\end{equation}}
\newcommand{\bea}{\begin{eqnarray}}
\newcommand{\eea}{\end{eqnarray}}

\def\ba{\begin{eqnarray}}
\def\ea{\end{eqnarray}}

\definecolor{darkblue}{RGB}{0,0,196}
\usepackage[colorlinks=true,linkcolor=darkblue,citecolor=darkblue,urlcolor=darkblue]{hyperref}

\begin{document}
\title{Design and development of advanced Al-Ti-V alloys for beampipe applications in particle accelerators}

\author{Kamaljeet Singh$^{1}$}
\author{Kangkan Goswami$^{1}$}
\author{Raghunath Sahoo$^{1}$}
\email[Corresponding Authors: ]{Raghunath.Sahoo@cern.ch}
\author{Sumanta Samal$^{2}$}
\email[]{sumanta@iiti.ac.in}

\affiliation{$^{1}$Department of Physics, Indian Institute of Technology Indore, Simrol, Indore 453552, India}
\affiliation{$^{2}$Department of Metallurgical Engineering and Materials Science, Indian Institute of Technology Indore, Simrol, Indore 453552, India}

\begin{abstract}
%In the context of heavy-ion collision experiments at the Large Hadron Collider (LHC) and Relativistic Heavy Ion Collider (RHIC), the beampipe is the first material seen by the produced particles. Therefore, it plays a crucial role in the signal-to-background ratio in subatomic particle detection. 

The present investigation reports the design and development of an advanced material with a high figure of merit (FoM) for beampipe applications in particle accelerators by bringing synergy between computational and experimental approaches. Machine learning algorithms have been used to predict the phase(s), low density, and high radiation length of the designed Al-Ti-V alloys. Al-Ti-V alloys with various compositions for single-phase and dual-phase mixtures, liquidus temperature, and density values are obtained using the Latin hypercube sampling method in TC Python Thermo-Calc software. The obtained dataset is utilized to train the machine-learning algorithms. Classification algorithms such as XGBoost and regression models such as Linear Regression and Random Forest regressor have been used to compute the number of phases, radiation length, and density respectively. The XGBoost algorithms show an accuracy of $98\%$, the Linear regression model shows an accuracy of $94\%$, and the Random Forest regressor model is accurate up to $99\%$.
The developed Al-Ti-V alloys exhibit high radiation length as well as a good combination of high elastic modulus and toughness due to the synergistic effect of the presence of hard $Al_3Ti$ phase along with a minor volume fraction of FCC $(Al)_{ss}$ solid solution phase mixture. The comparison of our alloys, alloy-1 ($Al_{75.2}Ti_{22.8}V_{2}$) and alloy-2 ($Al_{89}Ti_{10}V_{1}$) shows an increase in the radiation length by seven-times and a decrease in the density by two to three times as compared to stainless steel 304, the preferred material for constructing beampipes in low-energy particle accelerators. Further, we experimentally verify the elastic modulus of the alloy-1 and compute the FoM equal to 0.416, which is better than other existing materials for beampipes in low-energy experiments.
\end{abstract}
\date{\today}
\maketitle
\section{Introduction}
Beampipes are critical components in particle accelerators, serving as the vacuum chambers through which particle beams travel. The design and material selection of beampipes are influenced by various factors, including the type of particles being accelerated, their energy, and the specific applications of the accelerator. In high-energy physics, beampipes are used in facilities like the Large Hadron Collider (LHC) and Relativistic Heavy Ion Collider (RHIC) to accelerate protons and heavy ions to nearly the speed of light before colliding them to explore the characteristics of primordial matter the universe~\cite{Sahoo:2021aoy,Busza:2018rrf}. Beampipes passing through the four LHC experiments are the most intimate interface between machines and experiments. Therefore, beampipe is a critical component that lies at the heart of these experiments~\cite{art,Burkhardt2012}. The beampipe in these experiments is important for guiding particle beams through the accelerators and helps to maintain their trajectory. It prevents the beam particles from colliding with air molecules by providing them with a vacuum environment. For such experiments, one can define an ideal beampipe with a material completely transparent to radiation, holding an absolute vacuum to prevent any residual gas atoms or molecules from colliding with accelerated particles and having a small diameter so that created particles can be tracked from the interaction point~\cite{knaster,bruning2015high}. The basic requirement of materials for beampipes in particle accelerators is the minimum of interference with the particle detectors.  The required transparency of materials can be quantified either by calculating their radiation length ($X_0$), defined as the average length (in meter) in the material through which an electron loses its energy by the factor 1/e through electromagnetic interaction and multiple scattering~\cite{Gupta2010CalculationOR}. The radiation length decreases as the atomic number ($Z$) increases. A material with high $X_0$ is always desirable for designing the beampipe for particle accelerators working at any energy scale. The choice of higher $X_0$ minimizes unwanted interactions between the beam/produced particles and the beampipe material, preserving beam quality and enhancing the signal-to-noise ratio in the measurements. Materials like beryllium, which has a high radiation length, are often used as beampipes in high-energy experiments such as the RHIC and LHC. On the other hand, a low $X_0$ material would increase the background in the measurements due to an increase in unwanted interactions~\cite{Veness:2002ab}.  Moreover, a certain material thickness is required to resist the pressure loading for the chamber wall or bending stresses for a flange, which is quantified by the elastic modulus ($E$)~\cite{Veness:2011zz}. Some studies aim to fulfill several different physics goals that can be pursued by inserting mini-detectors inside the beampipe at large and small angles~\cite{Ayres:1981zz, Zunica:2023ped}. In collider experiments, beampipes serve multiple purposes beyond merely containing particle beams. They act as primary vacuum barriers, maintaining ultrahigh vacuum conditions to minimize particle scattering and energy loss due to collisions with residual gas molecules. The applications of beampipes are not only limited to particle accelerators but also include nuclear research and high-power hadron beam applications. Regarding the material of beampipes for any particle accelerators used in nuclear physics, the two main variables, atomic number and mass number ($A$) are considered to minimize the matter-particle interaction. The tendency of the material to slow down the energetic particles traversing through its interior is called stopping power ($S$)~\cite{CORREA2018291}. Since the stopping power scales with $Z$, low-$Z$ material is required for higher transparency to the energetic particles. 

In the early days of particle accelerators, beampipes were typically simple cylindrical tubes made of materials like stainless steel or copper. These pipes served primarily to guide and contain particle beams as they traversed through the accelerator complex. However, with advancements in accelerator technology and the emergence of high-energy collider experiments, the demands on beampipes grew more complex. As the beam energies increased, the choice of materials for beampipes became critical. Materials needed to withstand high vacuum conditions, intense radiation levels, and potential impacts from particle collisions. Stainless steel, beryllium, aluminum alloys, and titanium are commonly used due to their favorable mechanical properties and compatibility with ultra-high vacuum environments, along with their sensitivity towards high energy radiation and signal-to-background ratio.
The experiments at the LHC, such as A Large Ion Collider Experiment (ALICE)~\cite{ALICE:2022wwr} and A Toroidal LHC Apparatus (ATLAS)~\cite{ATLAS:2023dns}, use beryllium beampipes due to their low material budget and strength~\cite{art}. In the ALICE experiment, the current beampipe is about one meter long, 36.4 mm in diameter, compared to 50 mm before, and with a thickness of 0.8 mm, which is at the limit of what can be achieved with current technology. The other properties of beryllium, such as low elongation and high fragility, make it complicated to machine and practically unfeasible to weld because it gets cracks under thermal stress. The toxic nature of beryllium is also a major demerit in machining~\cite{knaster}. Although beryllium can be welded using advanced techniques such as Electron Beam welding, Laser Beam welding, etc, its inherent brittleness, toxicity, and the requirements for safe handling and specialized equipment make it challenging for widespread use in beam pipe applications. This toxicity necessitates strict safety protocols and proper handling procedures, increasing the cost and complexity of using beryllium. Furthermore, beryllium is a relatively expensive material compared to alternatives such as aluminum alloys or stainless steel~\cite{Brunet:1979tq}. The high cost of beryllium can be a limiting factor for large-scale collider experiments or upgrades. Also, beryllium is a very difficult metal to work with as it comes in the form of a powder that must be compressed at very high pressure to obtain a bar of metal that is then hollowed out. Only a few facilities in the world can produce such components from beryllium. Beyond high-energy physics, beampipes also have a crucial role in synchrotron light sources and free-electron lasers, where they guide electron beams used to generate intense, coherent X-rays for materials science, biology, and medical research. In these applications, the beam pipe design focuses on minimizing impedance and ensuring the stability of the electron beam to produce high-quality X-ray beams.
In India, synchrotron radiation sources like Indus-2~\cite{Deb_2013, Bhawalkar1998} with beam energies of a few MeV have a beampipe of aluminum alloy 5083 H321 with a thickness of $\approx $ 0.2 mm and the FoM ($X_{0}E^{1/3} \approx $ 0.374). 
%Residual magnetic interference is also a concern for stainless steel because the manufacturing process (e.g., welding or forming) could introduce slight magnetic fields in localized regions. This magnetic interference can affect the accuracy and stability of beam trajectories, particularly in experiments requiring high-precision measurements~\cite{inproceedings}. To overcome these demerits, aluminum alloy 5083 H321 was chosen as the material for beampipe in Indus-2~\cite{Deb_2013, Bhawalkar1998} because of its high thermal conductivity 
%$\approx 117 W/mK$ 
%with the FoM ($X_{0}E^{1/3} \approx $ 0.374). 
%On the other hand, the materials we target for the beampipe design with dual phase are non-toxic in nature, have higher thermal conductivity $\approx 159 W/mK$, the higher figure of merit ($X_{0}E^{1/3} \approx $) and cost-effective. 
The goal of our current study is to design and develop the beampipe for low-energy accelerators running at the energy up to the scale of the order of a few MeV to GeV with improved values of FoM. 

In this paper, we explore recent advancements in beampipe design and the development of novel materials with improved radiation length $X_0$ and elastic modulus $E$. 
Recent advancements in machine learning have enabled researchers to employ computational tools for the tailored design of alloys. Machine learning (ML) has been instrumental in pushing progress in a wide variety of materials, including metallic glasses, high-entropy alloys, shape-memory alloys, magnets, superalloys, catalysts, and structural materials~\cite{Hart:2021nat}. A significant contribution to this field is the introduction of common and advanced machine-learning models and algorithms used in alloy design. This approach emphasizes that the acquisition, utilization, and generation of effective data are key issues for the development of ML models and algorithms for future high entropy alloys (HEAs) design~\cite{Liu:2024md}. In addition, Paturi et al. conducted a comprehensive review of the role of machine learning in the additive manufacturing of titanium alloys. They highlighted the use of machine learning algorithms such as artificial neural network (ANN), support vector machine (SVM), convolutional neural network (CNN), decision tree (DT), k-nearest neighbor (KNN), k-means clustering, random forest (RF), Bayesian networks, self-organizing maps (SOM), and Gaussian process regression (GPR) in the design, fabrication, development, and quality control of titanium components via additive manufacturing~\cite{Paturi:2023ac}. We take advantage of various classification and regression algorithms to prepare aluminum-based alloys that meet the specifications needed for beampipe fabrication. Al-based alloys have been a great sensation due to their unique structure and unusual properties, such as high specific strength, low density, good oxidation and corrosion resistance, and high thermal conductivity~\cite{MUNDHRA2024174288,NAYAK2010128}. The materials with high thermal conductivity values have a tendency to spread the continuously deposited heat~\cite{BAUER2001505}. Extensive research is being pursued on different types of Al-based alloys with unique microstructural features to obtain desired properties. Light-weight materials have evoked the interest of the scientific community due to their worthwhile ability to construct beampipes with high radiation length and elastic modulus. To the best of the authors’ knowledge, limited studies on the development of beampipe materials have been reported. However, the primary objectives of the current investigation, which are to design and develop advanced material with high radiation length and elastic modulus, are organized in the following manner.

First, an overview of the methods used to calculate the radiation length and elastic modulus values is provided in section (\ref{Sec-formalism}). In section (\ref{Sec-method}), the machine learning methods and thermodynamic simulations using Thermo-Calc software are reported. Then, the results of the current study are discussed in section (\ref{Sec-results}) and compared with previously existing data. Finally, in the section (\ref{sec-summary}), some conclusions from the current study are summarized with possible outlooks. 
\begin{figure*}
    \centering
    \includegraphics[width=0.445\linewidth]{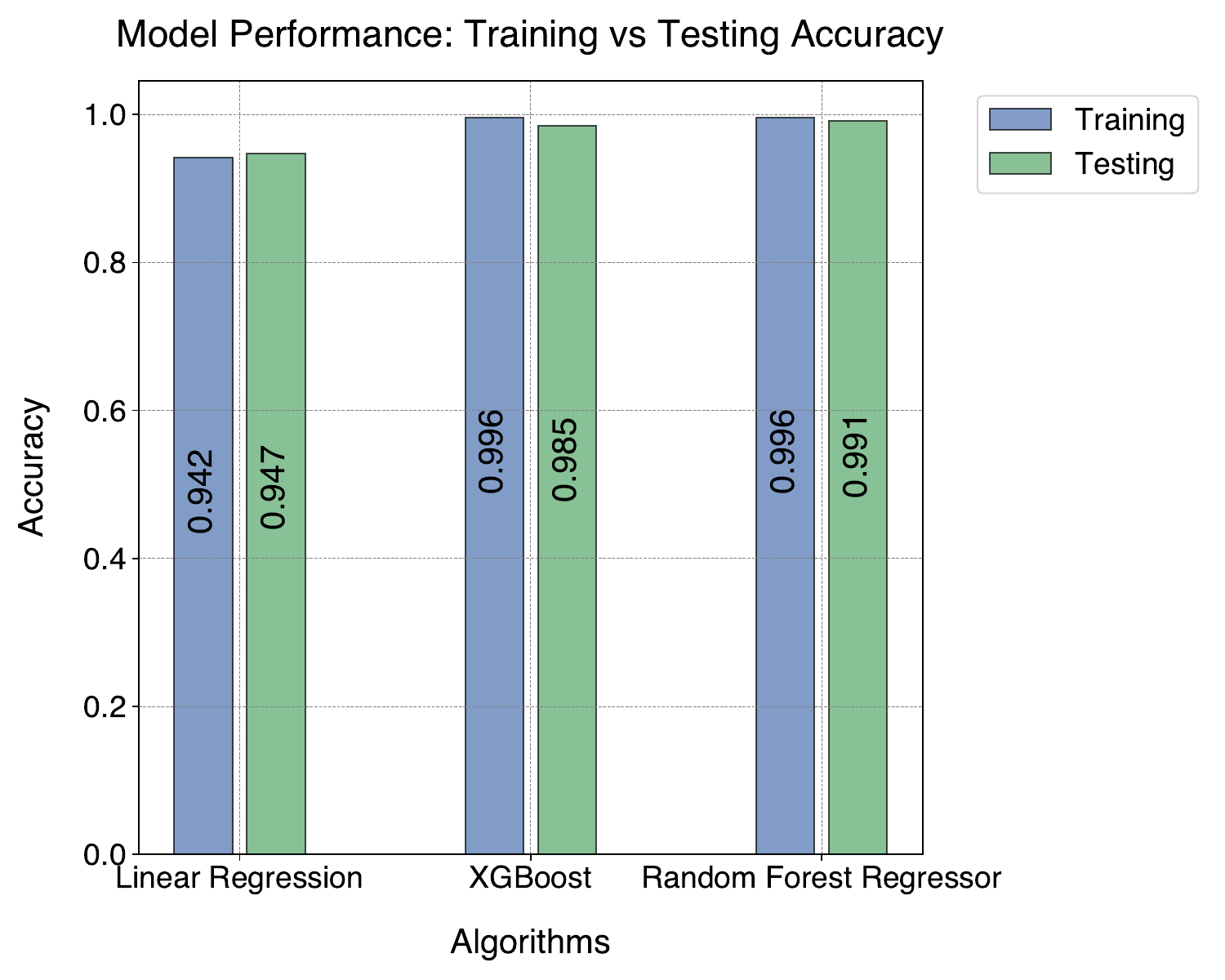}
    \includegraphics[width=0.4\linewidth]{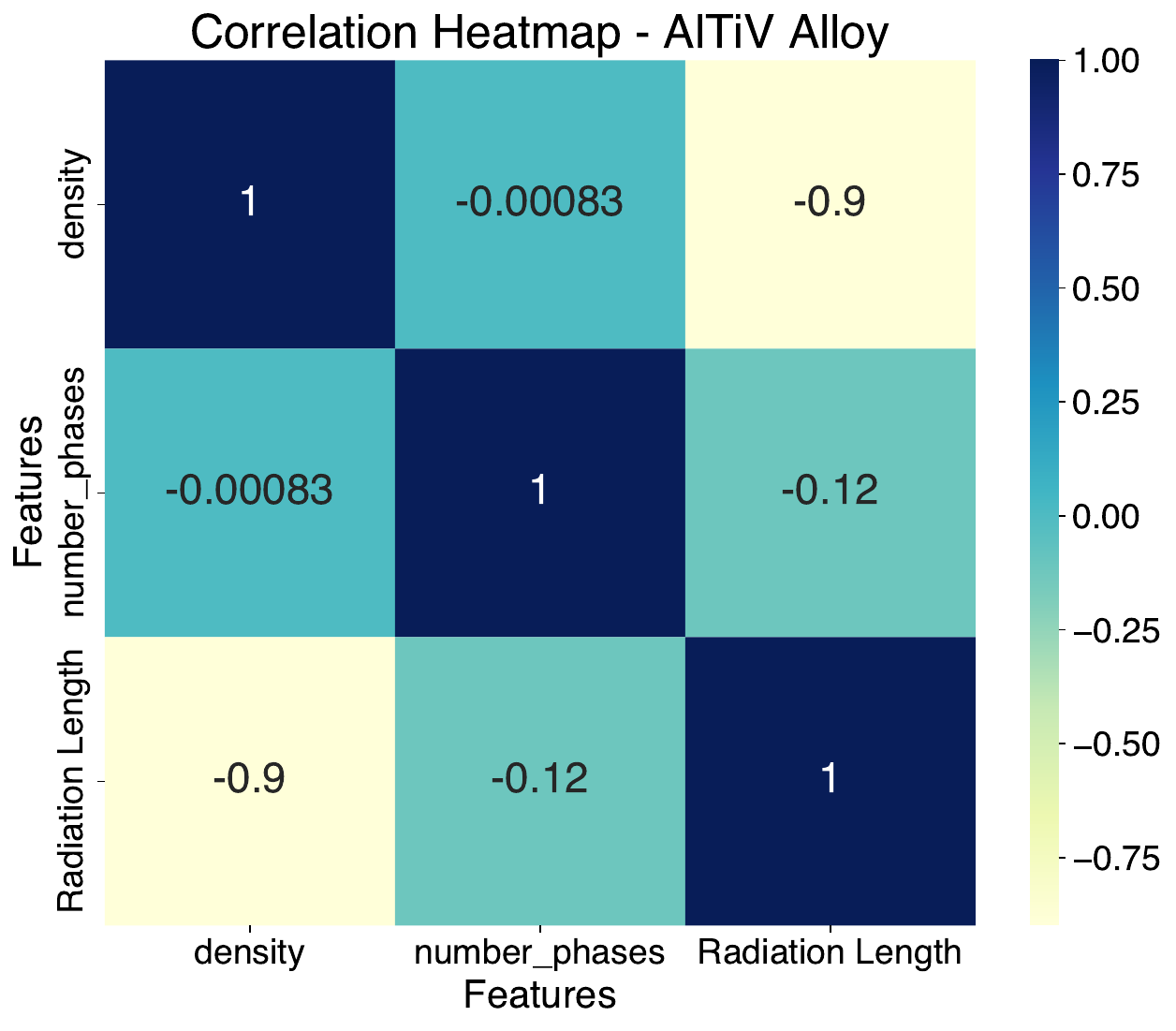}
    \caption{(Left Panel): Accuracies of three models. (Right Panel): Correlation matrix between the input feature in machine-learning models for AlTiV.}
    \label{fig:acc}
\end{figure*}
\section{Formalism}\label{Sec-formalism}
In this section, we discuss the formula used to calculate the radiation length values of the designed alloys. The experimentally measured values of the elastic modulus of designed materials are also discussed. 
\subsection{Calculation of radiation length ($X_0$)}
The radiation length ($X_0$) is a measure of the average distance traversed by an electron (or a charged particle) in the material over which the energy of the traversed particle is reduced by a factor of $1/e$ (around 37\%) due to electromagnetic interaction and multiple scattering. 
\begin{align}
    \epsilon = \epsilon_0 e^{-x/X_{0}},
\end{align}
where $\epsilon_0$ is the initial energy of the particle, and $x$ is the distance traversed in the material by an energetic particle.
We can calculate the radiation length constant ($kg/m^2$) as~\cite{Stolzenberg:2019ldl,PhysRevAccelBeams.27.024801}  
\begin{align}
    \Bar{X_0} = \frac{716.4 A}{Z(Z+1)\ln({287/\sqrt{Z}})}~(kg/m^2).
\end{align}
Where $Z$ is the atomic number, $A$ is the number of nucleons in an atom, and $\Bar{X_0}$ is the radiation length constant measured in $kg/m^{2}$. One can express radiation length ($X_0$) in units of length by dividing radiation length constant (${\Bar{X_0}}$) with the mass density $(\rho)$ of a material ($kg/m^3$):
\begin{align}
 X_0 = \frac{\Bar{X_0}}{\rho}~(m).   
\end{align}
\subsection{Experimental measurements of elastic modulus ($E$)}
 The room-temperature nanoindentation test has been carried out to evaluate the mechanical properties of alloy-1 ($Al_{75.2}Ti_{22.8}V_{2}$). The elastic modulus ($E$) of alloy-1 is estimated using volume fraction of $(Al)_{ss}$ and $Al_{3}Ti$ phases. The experimentally measured value of $E$ is found to be 96.9 and 173.4 GPa for $(Al)_{ss}$ and $Al_{3}Ti$ phases, respectively. The corresponding FoM of alloy-1 is given in Table.~\ref{tab:comp}.

\section{Methodology}\label{Sec-method}
First, we used the TC Python script to extract the data of various different compositions to design the required alloys. In Thermo-Calc, TC Python is a scripting language specifically designed for performing thermodynamic calculations and accessing the functionality of the Thermo-Calc software. TC Python provides access to the extensive thermodynamic databases available in Thermo-Calc. Users can specify the phases and components involved in their calculations and the thermodynamic models and parameters to be used. Further, we use this extracted data to train the machine-learning models and find the best-required composition in terms of the highest value of radiation length and number of phases. The industrially processed alloys follow the non-equilibrium solidification pathways due to solute segregation. Therefore, we follow the Gulliver–Scheil model which is generally used as the appropriate model for predicting qualitatively solidification pathways because of the limited solute diffusion in the solid. This model describes the solute redistribution during solidification by assuming complete diffusion in liquid, with no diffusion in solid and local equilibrium at the interface. The liquid concentration ($C_L$) at solid-liquid ($S/L$) interface can be expressed as~\cite{samal2016solidification,zuo2013evolution}:
\begin{align}
   C_L = C_0 (1 - f_s)^k-1 
\end{align}
where $C_0 $ is the initial liquid concentration, $k$ is the equilibrium partition coefficient, and $f_s$ is the solid fraction.

\subsection{Machine learning models}

Machine learning excels in solving classification and regression problems due to its ability to learn from data and make predictions without being explicitly programmed. For classification problems, machine learning algorithms can efficiently categorize unseen data into predefined classes based on patterns learned from training data. On the other hand, for regression problems, machine learning can predict continuous outcomes by learning the relationship between input features and the target variable. The efficiency of machine learning in these areas originates from its ability to handle large datasets and identify correlations among the input and target variables, making it a powerful tool for both classification and regression tasks. 

\begin{table}[h!]
    \centering
    \begin{tabular}{|c|c|}
        \hline
        \textbf{Hyperparameter} & \textbf{Default Value} \\
        \hline \hline
        fit\_intercept          & \texttt{True} \\
        \hline
        n\_jobs                 & -1 \\
        \hline
        positive                & \texttt{False} \\
        \hline
    \end{tabular}
    
    \caption{Hyperparameters of the Linear Regression model.}
    \label{tab:linear-regression-hyperparameters}
\end{table}

\begin{table}[h!]
    \centering
    \begin{tabular}{|c|c|}
        \hline
        \textbf{Hyperparameter} & \textbf{Default Value} \\
        \hline \hline
        booster             & \texttt{gbtree} \\
        \hline
        learning\_rate      & \texttt{0.3} \\
        \hline
        max\_depth          & \texttt{6} \\
        \hline
        subsample           & \texttt{1} \\
        \hline
        objective           & \texttt{binary:logistic} \\
        \hline
        n\_estimators       & \texttt{100} \\
        \hline
    \end{tabular}
    
    \caption{Hyperparameters of the XGBoost Classifier model.}
    \label{tab:xgboost-hyperparameters}
\end{table}

\begin{table}[h!]
    \centering
    \begin{tabular}{|c|c|}
        \hline
        \textbf{Hyperparameter} & \textbf{Default Value} \\
        \hline \hline
        n\_estimators           & \texttt{100} \\
        \hline
        criterion               & \texttt{squared\_error} \\
        \hline
        min\_samples\_split      & \texttt{2} \\
        \hline
        min\_samples\_leaf       & \texttt{1} \\
        \hline
        max\_features           & \texttt{sqrt} \\
        \hline
    \end{tabular}
    
    \caption{Hyperparameters of the RandomForest Regressor model.}
    \label{tab:randomforest-hyperparameters}
\end{table}

\subsubsection{Input to the machine learning algorithms}
\begin{figure}
    \centering
    \includegraphics[width= 0.8\linewidth]{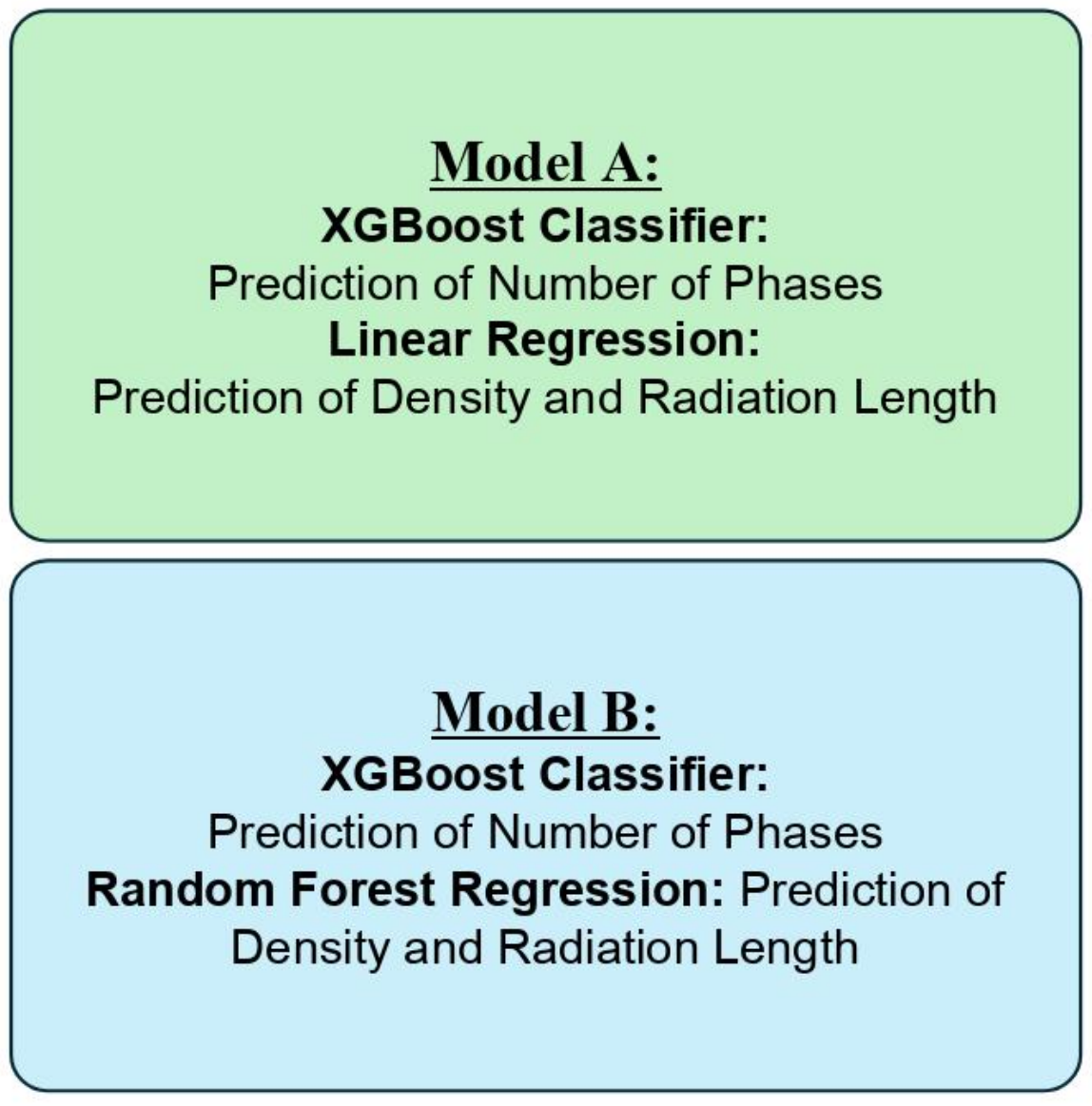}
    \caption{A visual representation of the Machine Learning models A and B.}
    \label{fig:Model}
\end{figure}

In this study, we train the machine learning models with 3 input features, radiation length, density, and number of phases, obtained for a specific composition of alloy using Thermo-Calc. We apply two different models: Model A, which combines an XGBoost Classifier~\cite{Chen:2016, readthedocs} for phase prediction and Linear Regression for predicting density and radiation length, and Model B, which combines an XGBoost Classifier with a Random Forest Regressor~\cite{Louppe:2014} for the prediction of density and radiation length, visually depicted in Fig.~\ref{fig:Model}. 

The input data from Thermo-Calc is split into 80 $\%$ for training and 20$\%$ for testing. We take a dataset of 2000 samples and their properties estimated using Thermocalc.
In Fig.~\ref{fig:acc}, we plot the accuracies of the individual ML algorithms. We observe that the XGBoost Classifier and Random Forest regressor perform better than the linear regression model, which is expected as linear regression is a very simple algorithm. Due to the high accuracy achieved by the XGBoost Classifier and Random Forest Regressor models, we have chosen to use their default hyperparameters, tabulated in Table~\ref{tab:linear-regression-hyperparameters}, \ref{tab:xgboost-hyperparameters}, and \ref{tab:randomforest-hyperparameters}, respectively. Moreover, changing the hyperparameters resulted in a very small improvement in the accuracy, but it resulted in substantially higher computational costs.
Furthermore, to study the correlation among the input features, we plot the correlation matrix, shown in the right panel of Fig.~\ref{fig:acc}. One can directly observe a strong negative correlation between density and radiation length for both of the alloys. However, there is a very small correlation between density and number of phases.

\subsubsection{Machine learning prediction}

\begin{figure}
    \centering
    \includegraphics[width=1\linewidth]{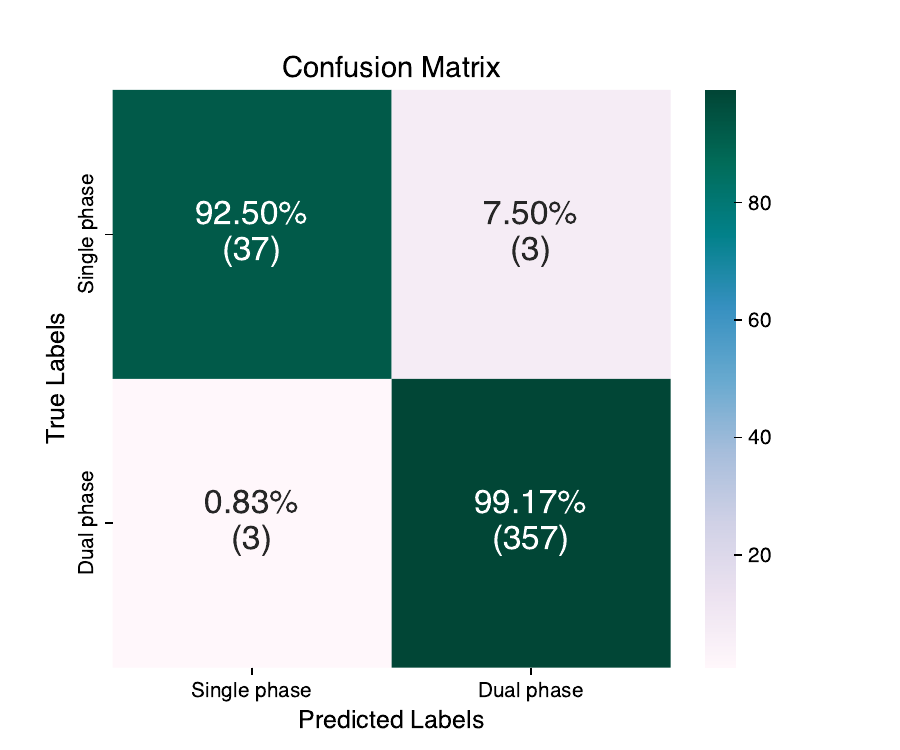}
    \caption{Confusion matrix for the classification of the phases using XGBoost model}
    \label{fig:confusion_matrix}
\end{figure}

We train two machine learning algorithms with different regressor models and use them to predict the properties of both single-phase and double-phase alloys composed of Al-Ti-V. These predictions are made using both Model A and Model B. To better understand the phase classification predictions made by the XGBoost model, we have plotted the confusion matrix in Fig.~\ref{fig:confusion_matrix}. We observe that the model successfully predicts the number of phases with higher accuracy; however, in our training and testing data, the number of single-phase alloys is much lower than the dual-phase alloys.
Moreover, we have tabulated the Precision, Recall, and F1-score of the model for the respective class. For single-phase alloys, the values for all of them are around 0.93, while for dual-phase alloys, the values are near 0.99.
In Table~\ref{tab:tab1}, we present the predicted composition of the Al-Ti-V alloy for a single-phase alloy, along with the corresponding number of phases, radiation length, and density. Similarly, Table~\ref{tab:tab2} contains the predictions for the dual-phase Al-Ti-V. In Table~\ref{tab:comp}, we compare the properties of the alloys predicted by the ML algorithm with Stainless steel 304, which is one of the preferred materials for the construction of beampipe for the low energy experiments.
\begin{table}[h!]
    \centering
    \begin{tabular}{|c|c|c|c|}
        \hline
        \textbf{Class}       & \textbf{Precision} & \textbf{Recall} & \textbf{F1-Score} \\
        \hline \hline
        Single phase         & \texttt{0.93}      & \texttt{0.93}    & \texttt{0.93}     \\
        \hline
        Dual phase           & \texttt{0.99}      & \texttt{0.99}    & \texttt{0.99}     \\
        \hline
    \end{tabular}
    \caption{Precision, Recall, and F1-Score for each class.}
    \label{tab:precision-recall-f1}
\end{table}
\begin{table}
    \centering
    \begin{tabular}{|c|c|c|}
        \hline
        & Model A & Model B \\
        \hline \hline
        Al & 0.752 & 0.752 \\
        \hline
        Ti & 0.228 & 0.159\\
        \hline
        V & 0.020 & 0.089\\
        \hline
        No. of phases & 1 & 1\\
        \hline
        $X_{0}~(m)$ & 0.0756 & 0.0749\\
        \hline
        Density ($kg/m^3)$& 3255.71 & 3361.21\\
        \hline
    \end{tabular}
    
    \caption{Model predictions for the compound Al-Ti-V and their properties for a single-phase alloy.}
    \label{tab:tab1}
\end{table}

\begin{table}
    \centering
    \begin{tabular}{|c|c|c|}
        \hline
        & Model A & Model B \\
        \hline \hline
        Al & 0.890 & 0.888 \\
        \hline
        Ti & 0.100 & 0.101\\
        \hline
        V & 0.010 & 0.011\\
        \hline
        No. of phases & 2 & 2\\
        \hline
        $X_{0}~(m)$ & 0.0830 & 0.0828\\
        \hline
        Density ($kg/m^3)$& 2712.47 & 2744.98\\
        \hline
    \end{tabular}
    
    \caption{Model predictions for the compound Al-Ti-V and their properties for a dual-phase alloy.}
    \label{tab:tab2}
\end{table}

\begin{table*}
    \centering
    \begin{tabular}{|c|c|c|c|c|c|}
    \hline
       Material  & Radiation  & Elastic  & $X_{0}E^{1/3}$ & Density  & Melting  \\
         & Length $(X_{0})$ [m] & Modulus (E) [GPa] &  & [$kg/m^3$]& Temperature [$\rm ^{\circ}C$]\\
    \hline \hline
        Stainless Steel 304 & 0.0176 & 193 & 0.1017 & 7930 & 1400-1450\\
    \hline
    Indus-2~\cite{Deb_2013}&  0.09   &  72  & 0.374 & 2650 & 570 \\
    \hline
       Designed alloy-1 & 0.0756 & 166.2 & 0.416 & 3259 & 1402 \\
    %\hline
       % $\rm Al_{89.0}Ti_{10}V_{1}$ & 0.07494 & - & - & 2.74 & 1588 \\
    \hline
    \end{tabular}
    \caption{Comparison of properties of different materials used for low-energy beampipes}
    \label{tab:comp}
\end{table*}

\section{Results and Discussion}\label{Sec-results}
\begin{figure*}
     \centering
     \begin{subfigure}[b]{0.43\textwidth}
         \centering
         \includegraphics[width=\textwidth]{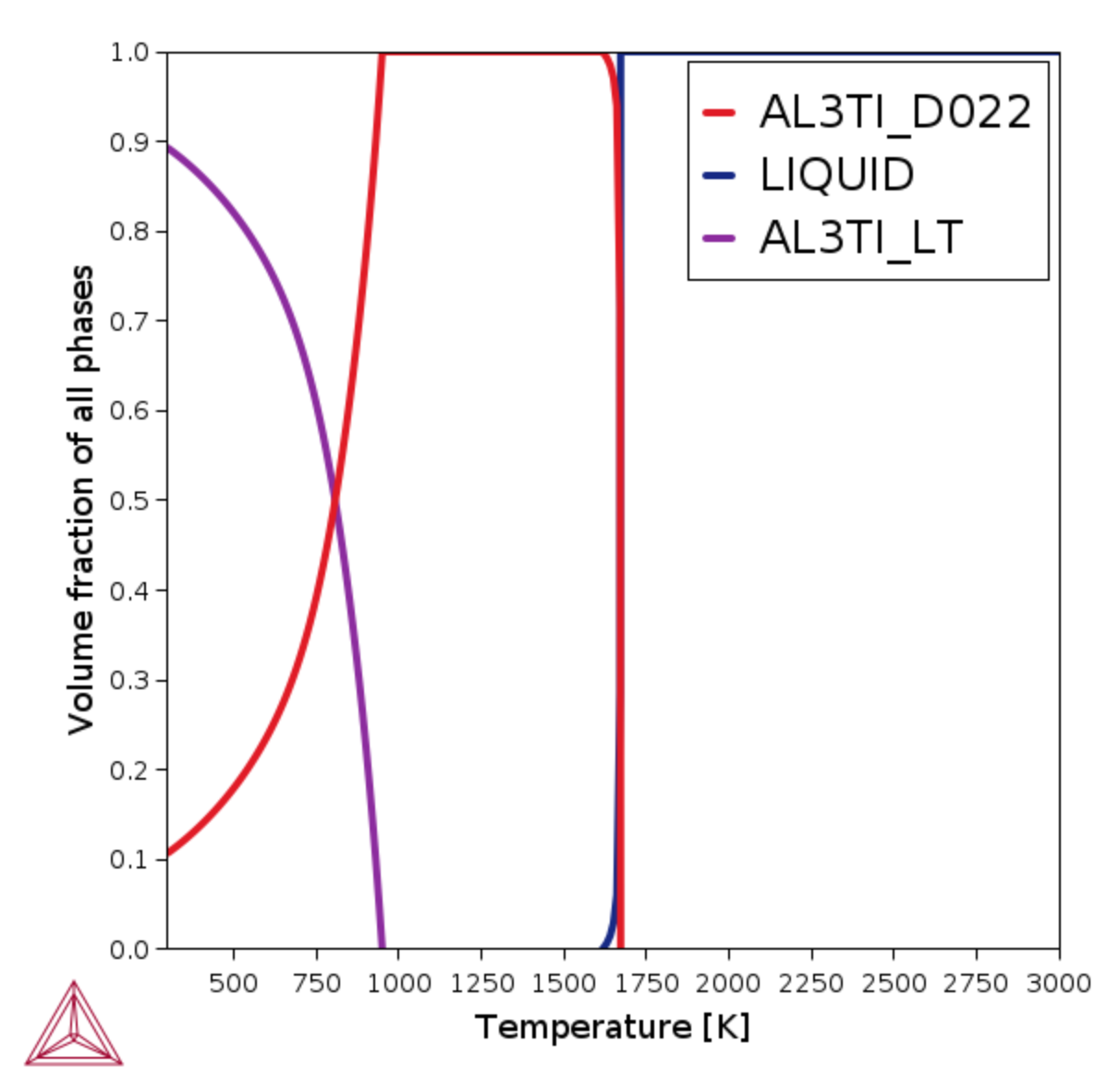}
          \caption{}
          \label{fig:svf}
     \end{subfigure}
     \hfill
     \begin{subfigure}[b]{0.43\textwidth}
         \centering
         \includegraphics[width=\textwidth]{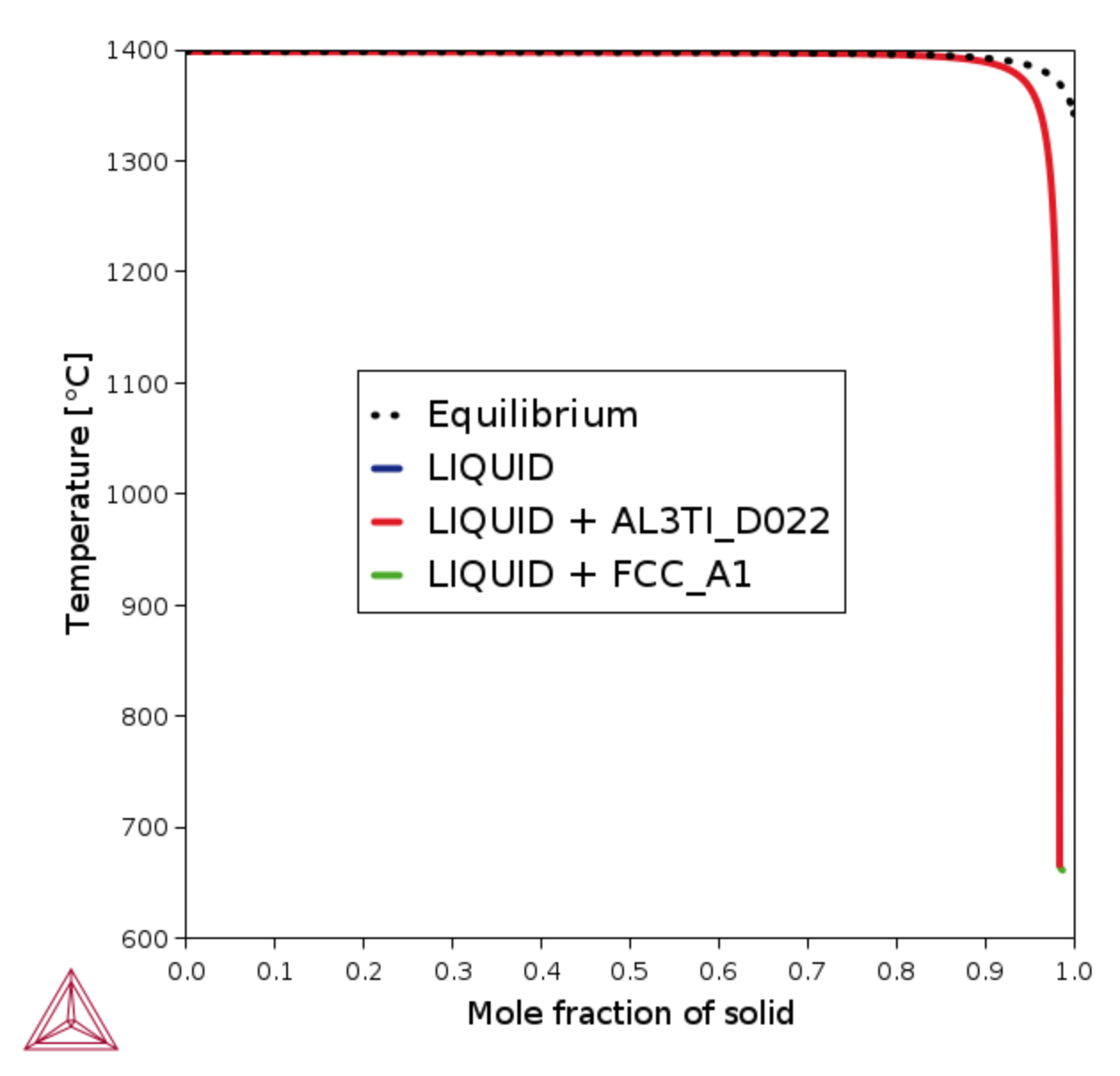}
          \caption{}
          \label{fig:smf}
     \end{subfigure}
     \hfill
     \begin{subfigure}[b]{0.43\textwidth}
         \centering
         \includegraphics[width=\textwidth]{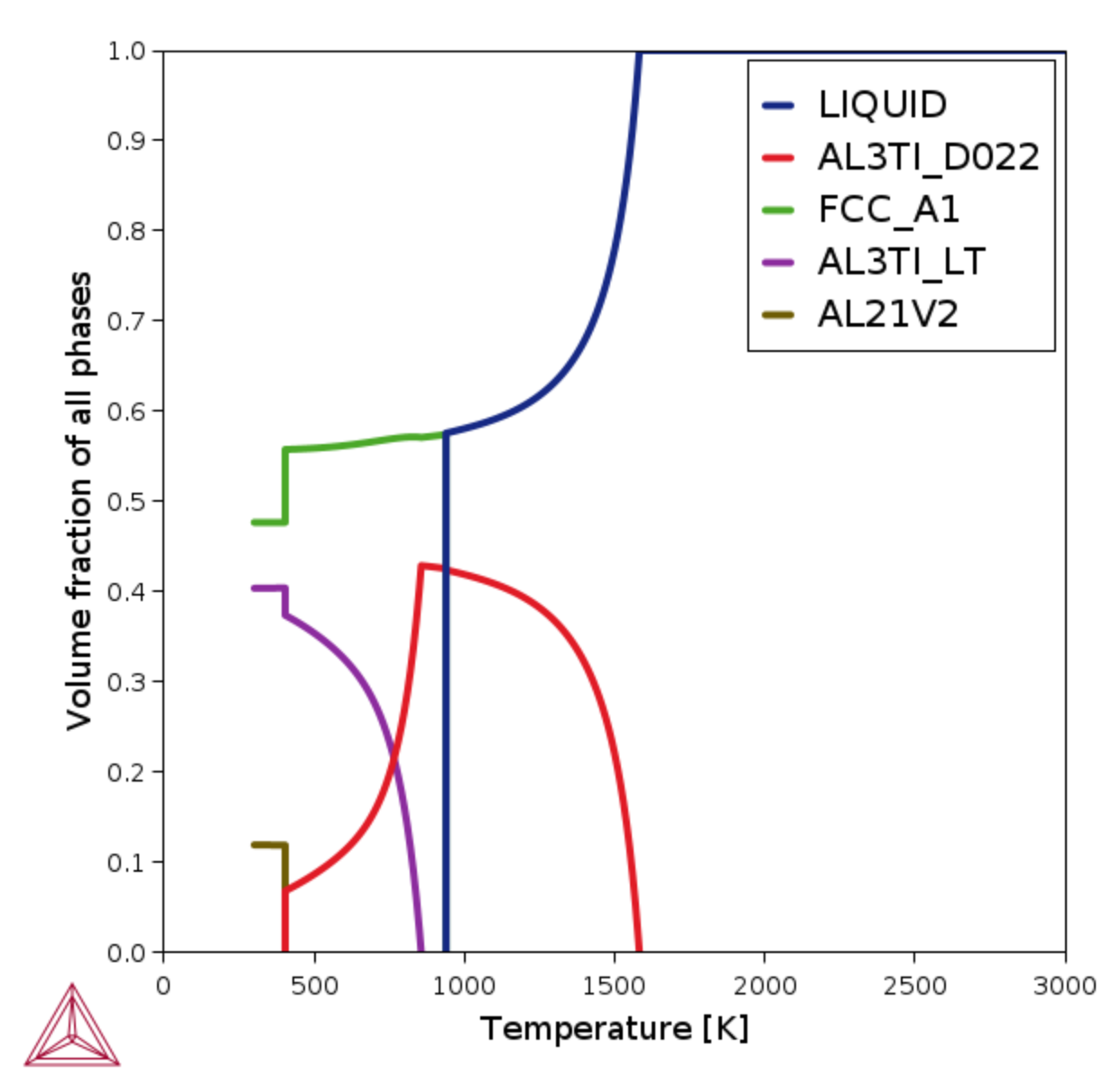}
          \caption{}
          \label{fig:dvf}
     \end{subfigure}
      \hfill
     \begin{subfigure}[b]{0.43\textwidth}
         \centering
         \includegraphics[width=\textwidth]{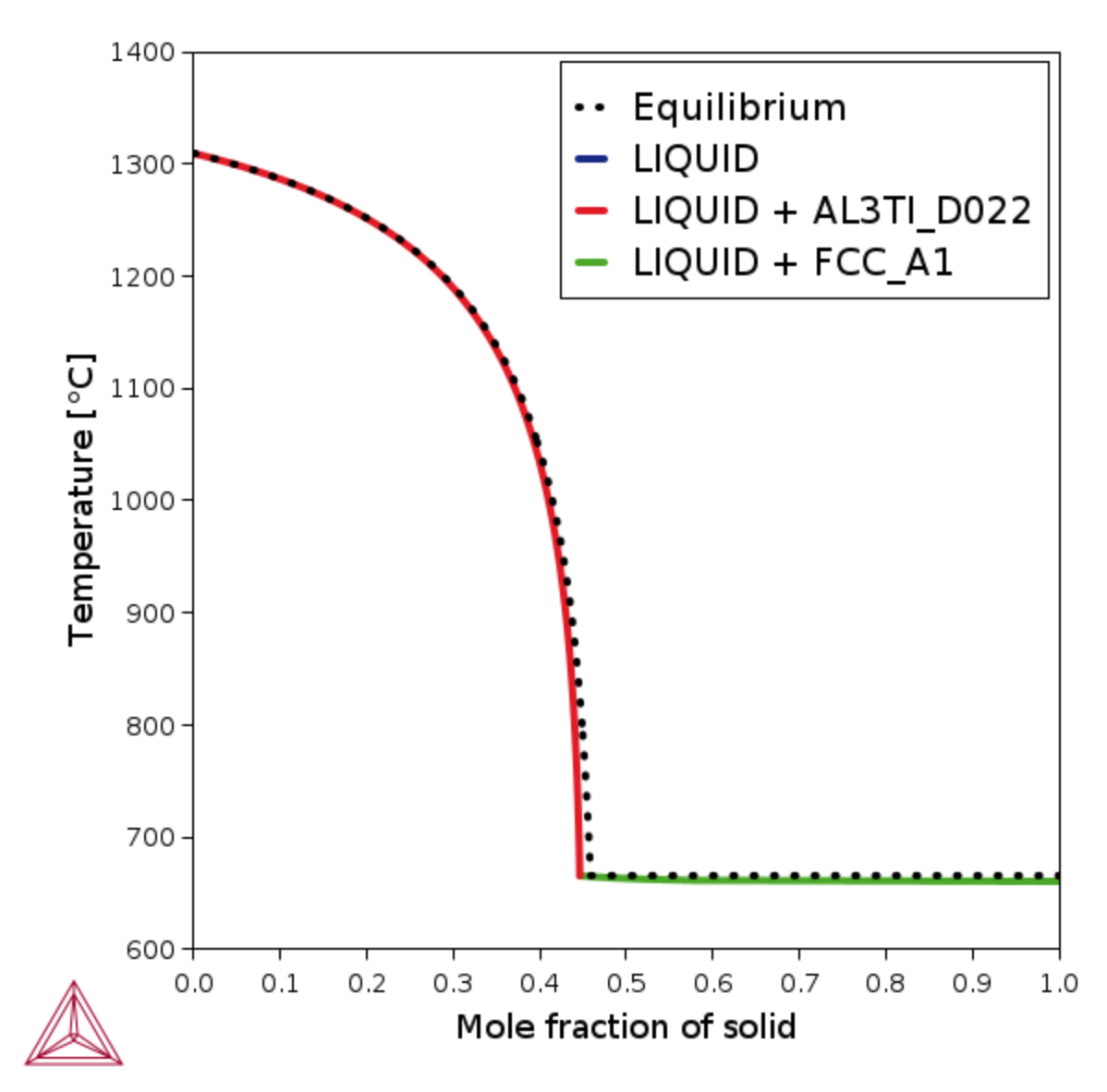}
          \caption{}
          \label{fig:dmf}
     \end{subfigure}
        \caption{Machine learning predictions for 
            (a) Volume fraction of solids with single-phase alloy. 
            (b) Mole fraction of solids with single-phase alloy. 
            (c) Volume fraction of solids with dual-phase alloy. (d) Mole fraction of solids with dual-phase alloy.}
	\label{Fig-2}
\end{figure*}

\begin{figure*}
     \centering
     \begin{subfigure}[b]{0.5\textwidth}
         \centering
         \includegraphics[width=\textwidth]{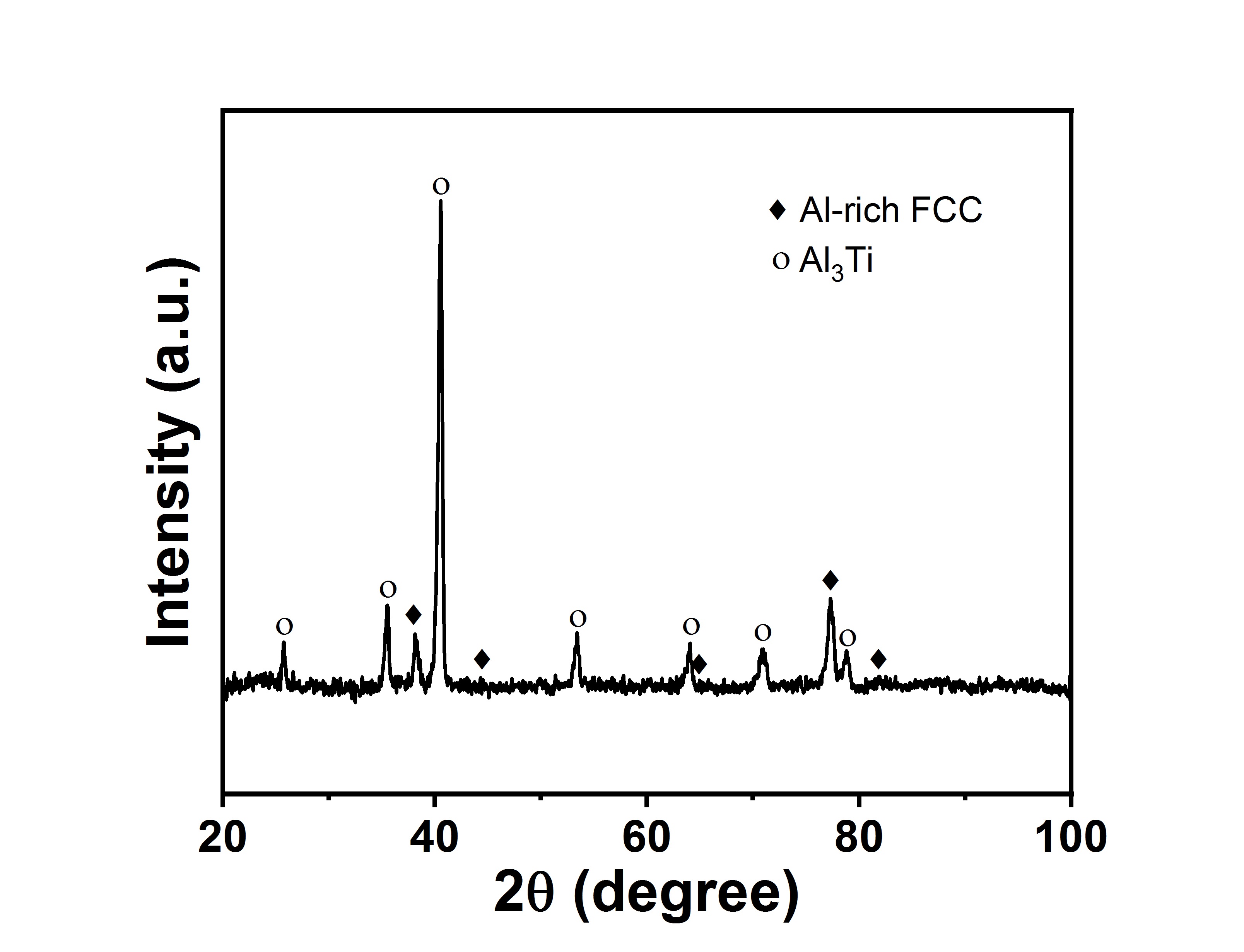}
          \caption{}
            \label{fig 3a}
     \end{subfigure}
     \hfill
     \begin{subfigure}[b]{0.43\textwidth}
         \centering
         \includegraphics[width=\textwidth]{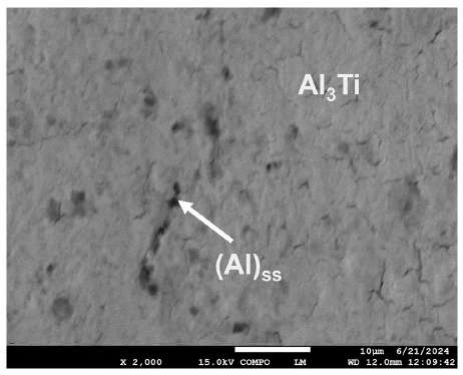}
          \caption{}
         \label{fig 3b}
     \end{subfigure}
     \hfill
     \begin{subfigure}[b]{1\textwidth}
         \centering
         \includegraphics[width=\textwidth]{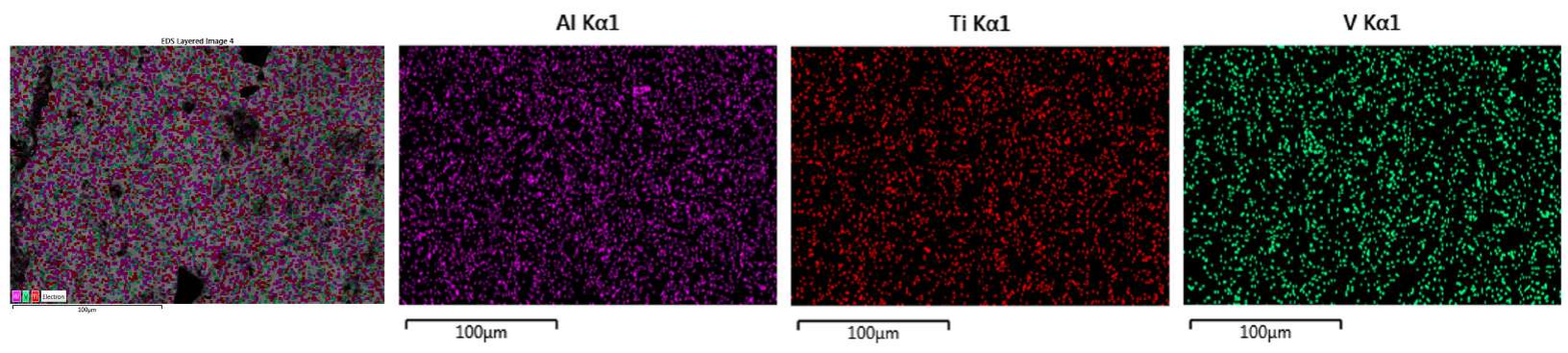}
          \caption{}
         \label{fig 3c}
     \end{subfigure}
      \hfill
        \caption{
            (a) XRD pattern, 
            (b) SEM micrograph, 
            (c) EDS mapping of alloy-1.} 
	\label{Fig-2}
\end{figure*}

\begin{figure*}
     \centering
     \begin{subfigure}[b]{0.5\textwidth}
         \centering
         \includegraphics[width=\textwidth]{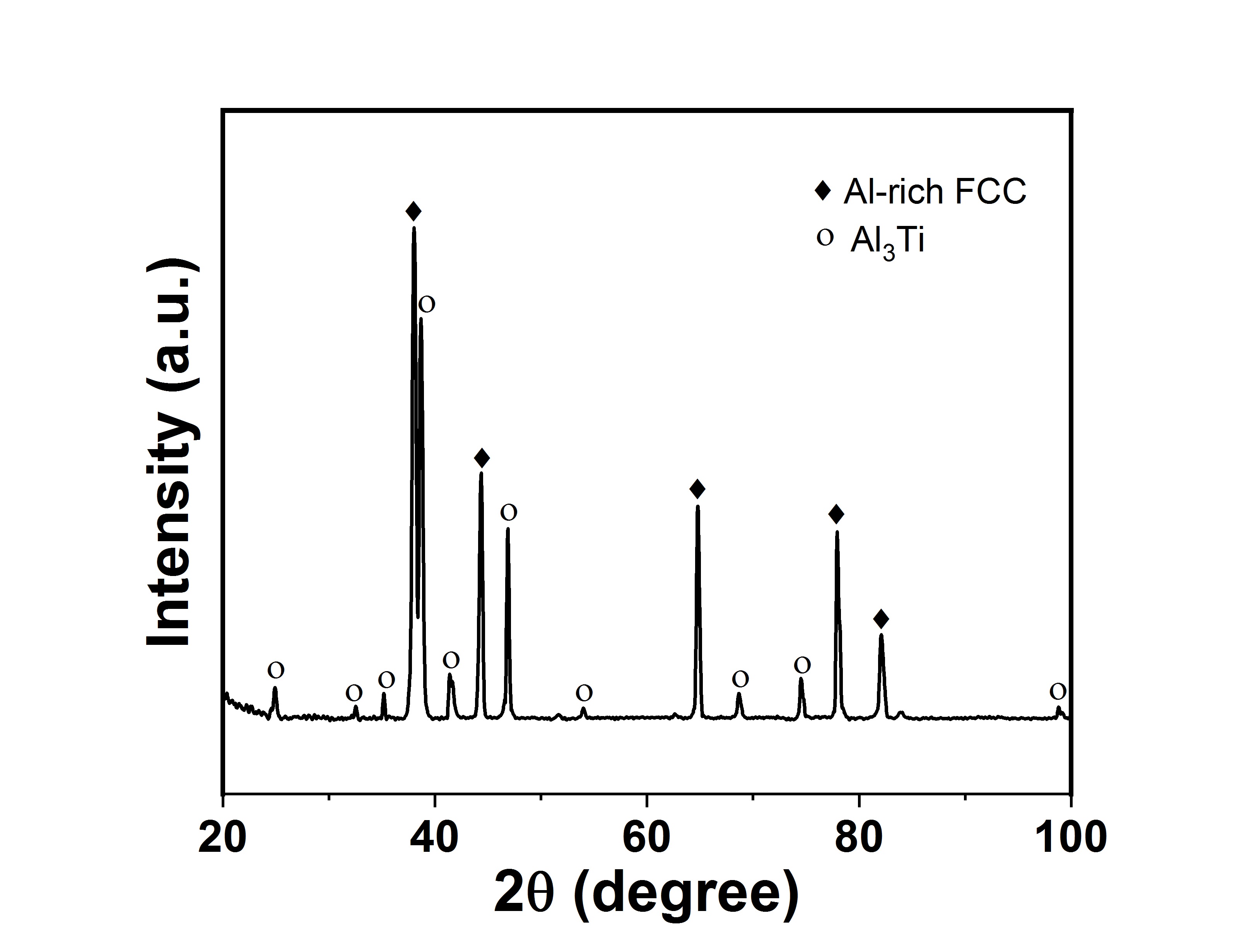}
          \caption{}
         \label{fig 4a}
     \end{subfigure}
     \hfill
     \begin{subfigure}[b]{0.43\textwidth}
         \centering
         \includegraphics[width=\textwidth]{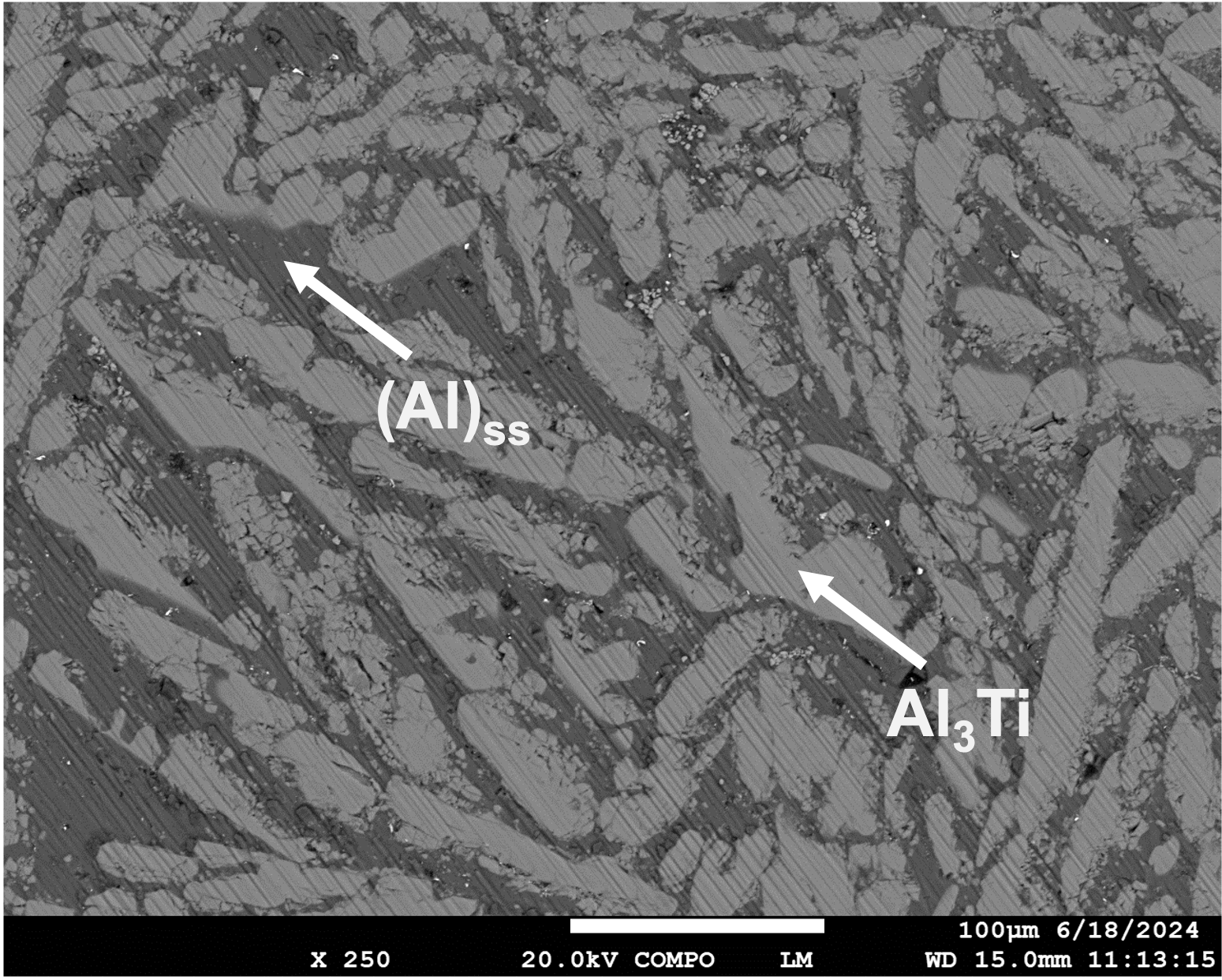}
          \caption{}
         \label{fig 4b}
     \end{subfigure}
     \hfill
     %\begin{subfigure}[b]{0.31\textwidth}
         %\centering
        %\includegraphics[width=\textwidth]%{DSC.pdf}
         % \caption{}
         %\label{fig 4c}
    % \end{subfigure}
     \hfill
     \begin{subfigure}[b]{1\textwidth}
         \centering
         \includegraphics[width=\textwidth]{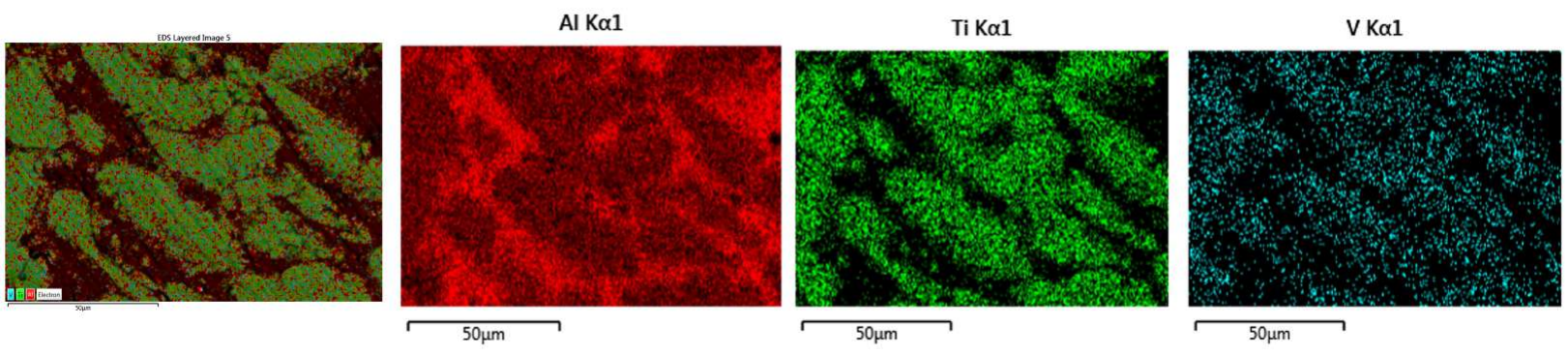}
          \caption{}
         \label{fig 4d}
     \end{subfigure}
      \hfill
       \caption{
            (a) XRD pattern, 
            (b) SEM micrograph,
            %(c) Thermal characterization using DSC
            (c) EDS mapping of alloy-2.} 
	\label{Fig-2}
\end{figure*}

\subsection{Comparison through Thermodynamic simulation}
The predicted phases of the designed Al-Ti-V alloys by ML technique are cross-validated with the thermodynamic simulation results. The thermodynamic simulation of alloy-1 has been performed using Thermo-Calc software (Version: Al database TCAL8). The volume fraction of phases Vs. temperature plot of alloy-1 is given in Figure~\ref{fig:svf}. Firstly, both FCC$\textunderscore$A1 and $Al_3Ti\textunderscore$D022 phases form from the liquid at high temperature, followed by phase transformation of FCC$\textunderscore$A1 to Al2V2 phase as well as $Al_3Ti\textunderscore$D022 to Al3Ti$\textunderscore$LT phase during solid-state transformation. Temperature Vs. mole fraction plot of Al-Ti-V alloy is given in Figure~\ref{fig:smf}, which shows the formation of $Al_3Ti\textunderscore$D022 phase (i.e. $L\rightarrow L+ Al_3Ti\textunderscore D022$) from the liquid, followed complete solidification of remaining liquid to form FCC$\textunderscore$A1 phase. It is found that the designed alloy exhibits almost single-phase alloy with a very high-volume fraction of $Al_3Ti\textunderscore$D022 phase (approx. 97-98 $\%$) and 2-3 volume fraction $\%$ of FCC$\textunderscore$A1 phase. Figure~\ref{fig:dvf} shows the volume fraction of phases Vs. temperature plot of alloy-2 ($Al_{89}Ti_{10}V_{1}$), highlighting the two-phase domain of $Al_3Ti\textunderscore$D022 is very large with liquidus temperature of ($\sim 1675$ K) and forms from the liquid at high temperature first. Subsequently, $Al_3Ti\textunderscore$D022 is transformed to $Al_3Ti\textunderscore$LT phase during solid-state transformation. Figure~\ref{fig:dmf} represents temperature Vs. mole fraction plot of $Al_{89}Ti_{10}V_{1}$ alloy, revealing the evolution of various phases during solidification. Firstly, $Al_3Ti\textunderscore$D022 phase is formed from the liquid as primary phase (i.e., $L\rightarrow L+ Al_3Ti\textunderscore D022$), followed by the formation of FCC$\textunderscore$A1 phase from the remaining liquid (i.e., $L\rightarrow L+ FCC\textunderscore A1 \rightarrow FCC\textunderscore A1$). The Al-Ti-V alloys consist of composite phases having both $Al_3Ti\textunderscore$D022 (approx. 57 $\%$) and FCC$\textunderscore$A1 ($\simeq43 \%$). The liquidus temperature for the studied alloy is calculated to be around $1590$ K. It is worth mentioning that the predicted results using ML techniques corroborate with the thermodynamic simulation outcomes.

\subsection{Experimental validation}
The designed Al-Ti-V alloys are synthesized by vacuum arc melting technique under a protective argon atmosphere in water-chilled Cu-mould using constituent elements having greater than 99.9 $\%$ mass purity. The alloy button is remelted at least 5 times to obtain a homogeneous chemical sample. The arc melted sample is cut and mirror polished for subsequent characterization of prepared samples. The structural characterization and microstructural characterizations of the developed Al-Ti-V alloys are carried out using an X-ray diffractometer (Model No.:Panalytical Empyrean X-ray Diffractometer) and Scanning electron microscopy (SEM, model no.:JEOL JSM-7610 Fplus model). The XRD pattern of alloy-1 (as shown in Figure \ref{fig 3a}) shows the presence of an intense peak of the $Al_3Ti$ phase and FCC $(Al)_{ss}$ phase. SEM coupled with EDS mapping results are given in Figures \ref{fig 3b} and \ref{fig 3c}, highlighting the presence of bright contrast $Al_3Ti$ phase and dark contrast Al-rich $(Al)_{ss}$ phase as well as uniform distribution of each element in the microstructure of the developed Al-Ti-V alloy. Figure \ref{fig 4a} shows the XRD pattern alloy-2, indicating the presence of FCC $(Al)_{ss}$ and $Al_3Ti$ phases. Figure \ref{fig 4b} reveals the SEM micrograph of alloy-2, showing a heterogeneous microstructure having an Al-rich $(Al)_{ss}$ phase (dark contrast) and $Al_3Ti$ phase (bright contrast). EDS mapping (as shown in Fig. \ref{fig 4d}) indicates the uniform distribution of each constituent element in the microstructure. Al-Ti-V alloy with a high volume fraction of $Al_3Ti$ intermetallics can be considered a potential candidate material for structural applications, particularly for low-energy beampipe applications. The elastic modulus of the individual phases of the developed alloy is 96.9 GPa for $(Al)_{ss}$ with a volume fraction of 9.4$\%$ and 173.4 GPa for $Al_3Ti$ with a volume fraction of 90.6$\%$, which makes the elastic modulus of the alloy-1 to be around 166.2 GPa.

\section{Summary}\label{sec-summary}
In this work, we design and develop a novel material for the application of beampipe in low-energy particle accelerators. The design of each beampipe is closely connected with the certain needs required by the environment in which it is installed. We deploy machine-learning algorithms to predict a suitable alloy of Al-Ti-V with the highest radiation length and lowest density. We make use of classifier algorithms (XGBoost Classifier) to predict the number of phases of the alloy and regressor models (Linear regression and Random Forest regressor) to predict the radiation length and density of the alloy. We generate data with the help of TC Python and use it for the training of the model. Upon validation of the model, we observe the accuracies of the models, Linear Regression, XGBoost, and Random Forest Regressor as $94\%$, $98\%$, and $99\%$, respectively. Theoretically, we predict two such alloys, one having a single phase and the other with two phases. We compare their properties using thermodynamic simulation results with Stainless steel 304, which is a preferred choice of material for the development of beampipe in low-energy particle accelerators. We find that the radiation length of our alloys is around seven times higher than that of the Stainless steel 304. The predicted alloys are relatively less dense, 3259 $kg/m^3$ and 2740 $kg/m^3$ for the alloy-1 ($Al_{75.2}Ti_{22.8}V_{2}$) and the alloy-2 ($Al_{89}Ti_{10}V_{1}$), respectively, as compared to the stainless steel 304, which has a density of 7930 $kg/m^3$. The material comparison suggests that the developed alloy-1 has excellent thermomechanical characteristics, better than other materials like stainless steel and other existing alloys because of their low density and excellent thermal conductivity. Further, we experimentally verify the elastic modulus of the alloy-1 and compute $X_{0}E^{1/3} \approx $ 0.416 and make the comparison of different materials as reported in table \ref{tab:comp}. We find that the FoM of our designed alloy-1 is better than the existing materials of the beampipe. Therefore, the developed material can be considered as a novel material for applications of beampipe in low-energy experiments. The reported Al-Ti-V alloys in the current work are targeted for the development and design of beampipes for low-energy accelerators. Light-weight multicomponent alloys will open up new vistas to discover novel materials with high radiation length and elastic modulus for beampipe applications in high-energy experiments such as the Large Hadron Collider and Relativistic Heavy Ion Collider.

\section*{Acknowledgments}
K.G. acknowledges the financial support from the Prime Minister's Research Fellowship (PMRF), Government of India. K.S. gratefully acknowledges the financial aid from UGC, Government of India. The authors acknowledge the DAE-DST, Government of India funding under the mega-science project “Indian participation in the ALICE experiment at CERN” bearing Project No. SR/MF/PS-02/2021-IITI(E-37123). The use of the computing facility at IIT Madras and useful discussions with Professor G. Phanikumar are highly appreciated by the authors. The nanoindentation facility of Dr. Sumeet Mishra, IIT Roorkee is thankfully acknowledged for the 
measurement of elastic modulus. The financial support provided by the Department of Science
and Technology (DST), Government of India, through the FIST funding
(SR/FST/ET-II/2021/778) is acknowledged.

%==================================================
\end{document}